\documentclass[sigconf,nonacm]{acmart}

\usepackage{booktabs}
\usepackage{array}
\usepackage{balance}
\usepackage{enumitem}
\usepackage{pifont}
\usepackage[ruled,vlined]{algorithm2e}
\usepackage{listings}
\usepackage{xcolor}
\definecolor{lstbg}{HTML}{F6F8FA}
\definecolor{lstkey}{HTML}{4338CA}
\definecolor{lststr}{HTML}{0F766E}
\definecolor{lstcom}{HTML}{64748B}
\lstdefinestyle{art}{
  basicstyle=\ttfamily\scriptsize, backgroundcolor=\color{lstbg},
  breaklines=true, breakatwhitespace=true, columns=fullflexible,
  keepspaces=true, showstringspaces=false, frame=leftline, framerule=1.2pt,
  rulecolor=\color{lstkey}, xleftmargin=6pt, framexleftmargin=6pt,
  aboveskip=4pt, belowskip=2pt, keywordstyle=\color{lstkey}\bfseries,
  stringstyle=\color{lststr}, commentstyle=\color{lstcom}\itshape,
  literate={\{}{{\{}}1 {\}}{{\}}}1 {[}{{[}}1 {]}{{]}}1,
}
\renewcommand\footnotetextcopyrightpermission[1]{}
\begin{document}

\title{Reconcile Once, Write Anytime: A Trust-Tiered Librarian and a
Multi-Agent Writer for Drift-Free, Point-in-Time Research}

\author{Xing Zhang, Yanwei Cui, Guanghui Wang, Peiyang He\textsuperscript{*}}
\affiliation{%
  \institution{AWS Generative AI Innovation Center}
  \country{}
}
\renewcommand{\shortauthors}{Zhang et al.}
\thanks{\textsuperscript{*}Corresponding author: \texttt{peiyan@amazon.com}.}

\begin{abstract}
Long-form research reports generated by large language models (LLMs) drift,
contradict themselves, and lose provenance: the same metric appears with different
values in different sections, numbers arrive without sources, and rumor is quoted
as confidently as an audited filing. We present a deployment-oriented two-tier
agentic system that separates a \emph{maintained, point-in-time knowledge library}
from \emph{report writing}. A deterministic ``librarian'' continuously ingests
public, timestamped sources into a trust-tiered \emph{ontology}, layering evidence
cards, an authoritative metric ledger, and a claim graph into an always-current
source of truth, not per-query RAG over raw chunks. A portable multi-agent
``writer'' runtime then composes a long, contradiction-free, evidence-grounded report
at any knowledge cutoff~$T$, reading only evidence with $\mathit{as\_of}\le T$ (no
look-ahead); red-team verdicts flow back into the librarian, closing the loop.
We evaluate on a self-collected, production-scale public corpus of \textbf{6{,}130
sources} yielding \textbf{555{,}926 evidence cards} (SEC EDGAR filings across 295
issuers and 11 sectors, U.S. Bureau of Labor Statistics macro releases, and
Wikipedia). From the \emph{one} library we compose four flagship point-in-time
reports on distinct theses (AI-compute, energy, healthcare/pharma, banks) and run
eight mechanical, reproducible experiments, whose headline metrics come from a
deterministic quality-control (QC) gate, itself validated by defect-injection
meta-evaluation at recall~$1.0$ \emph{and} precision~$1.0$ against negative controls.
A shared metric ledger removes 6{,}845 cross-section figure
contradictions to zero. Tier-first selection is correct on $22/22$ gold cases where a
popularity-first baseline scores only $9/22$; trust tiering leaks zero media-sourced
numbers into hard evidence, and no government statistic displaces a company's own
filing. A red-team refutation propagates back through a source override and
self-corrects a later run with zero manual edits. Point-in-time replay exhibits zero
look-ahead violations across seven cutoffs while the library grows from 235{,}373 to
555{,}312 cards. Finally, difficulty-tiered model routing with bounded parallelism
\emph{exceeds} the all-Opus quality ceiling on a graded score while running
$3.7\times$ faster than serial, with cost and latency recorded on every run.
\end{abstract}


\maketitle

\section{Introduction}
Automated long-form research (equity notes, market landscapes, due-diligence
memos) is a natural target for LLM agents, yet naive generation fails in ways that
matter precisely where the stakes are highest. Three failure modes recur. (i)
\emph{Numeric drift}: a report cites a company's capital expenditure as one figure
in the revenue section and a different figure in the capital-intensity section,
because each paragraph was grounded independently. (ii) \emph{Provenance loss}:
numbers appear with no traceable source, and the reader cannot tell an audited
disclosure from a fluent hallucination~\cite{ji2023hallucination}. (iii)
\emph{Trust flattening}: a pre-launch rumor and a filed 10-Q are quoted with equal
confidence, because the system has no notion of source authority.

Retrieval-augmented generation (RAG)~\cite{lewis2020rag,gao2023ragsurvey} re-retrieves
raw chunks per query and grounds each answer in isolation, maintaining no
\emph{consistent, evolving} body of judgements, so contradictions and stale figures
recur at every generation; agentic writers \cite{wu2023autogen,hong2023metagpt}
coordinate sections but inherit the same gaps. We argue the fix is architectural:
reconcile evidence \emph{once} into a maintained ontology (trust-ranked, timestamped,
incrementally updated), separated from the \emph{writing} of any
report, so consistency and provenance are properties of the store, governed by source
authority rather than popularity.

We built such a system, validated it internally on a private corpus, and evaluate
it here on a \emph{production-scale} public corpus (6{,}130 sources and
555{,}926 evidence cards across 295 issuers and 11 sectors, from which we compose
four flagship reports on distinct theses), as an \textbf{industry case study with
end-to-end reproducible results}, not a state-of-the-art (SOTA) claim. Our contributions:
\begin{enumerate}[leftmargin=1.3em]
\item \textbf{A coupled two-phase system} (\S\ref{sec:system}): a deterministic
librarian maintains a timestamped knowledge ontology, so new filings incrementally
refresh the source of truth without re-indexing (Phase~A); and a portable, model-routed multi-agent runtime generates a point-in-time
report at cutoff~$T$ (Phase~B).
\item \textbf{A distributed multi-agent writer with a shared-store coordination
substrate} (\S\ref{sec:distributed}): heterogeneous agents (per-section composers
and an adversarial red-team prosecutor) run under bounded-concurrency parallelism
and difficulty-tiered model routing, coordinated \emph{indirectly} through the
trust-tiered store rather than by direct message passing, so concurrency never
costs cross-section consistency.
\item \textbf{Trust-tiered consistency mechanisms as deployed governance}
(\S\ref{sec:trust}): an official-first metric ledger that ranks source
\emph{authority} above popularity (tier~$\to$~corroboration~$\to$~recency), a typed
claim graph, and a six-check deterministic QC gate that \emph{blocks} delivery and is
itself meta-validated by defect injection (E5), not a post-hoc score.
\item \textbf{A self-correcting write-back loop} (\S\ref{sec:system}): a report-side
red-team refutation flows back as a source override that self-corrects the ledger at
the next cutoff with no manual edits, and idempotent regeneration flags any promoted
claim whose approved evidence silently vanishes (an ``anchor swap'') for
re-validation, so the library only ever improves.
\item \textbf{A mechanical, meta-eval-validated evaluation} (\S\ref{sec:eval})
others can reuse, including the distributed-execution cost/latency of routing and
bounded parallelism (recorded on every run) and a \emph{temporal point-in-time}
result the living library uniquely enables.
\end{enumerate}

\section{System}
\label{sec:system}
\begin{figure*}[t]
  \centering
  \includegraphics[width=\textwidth]{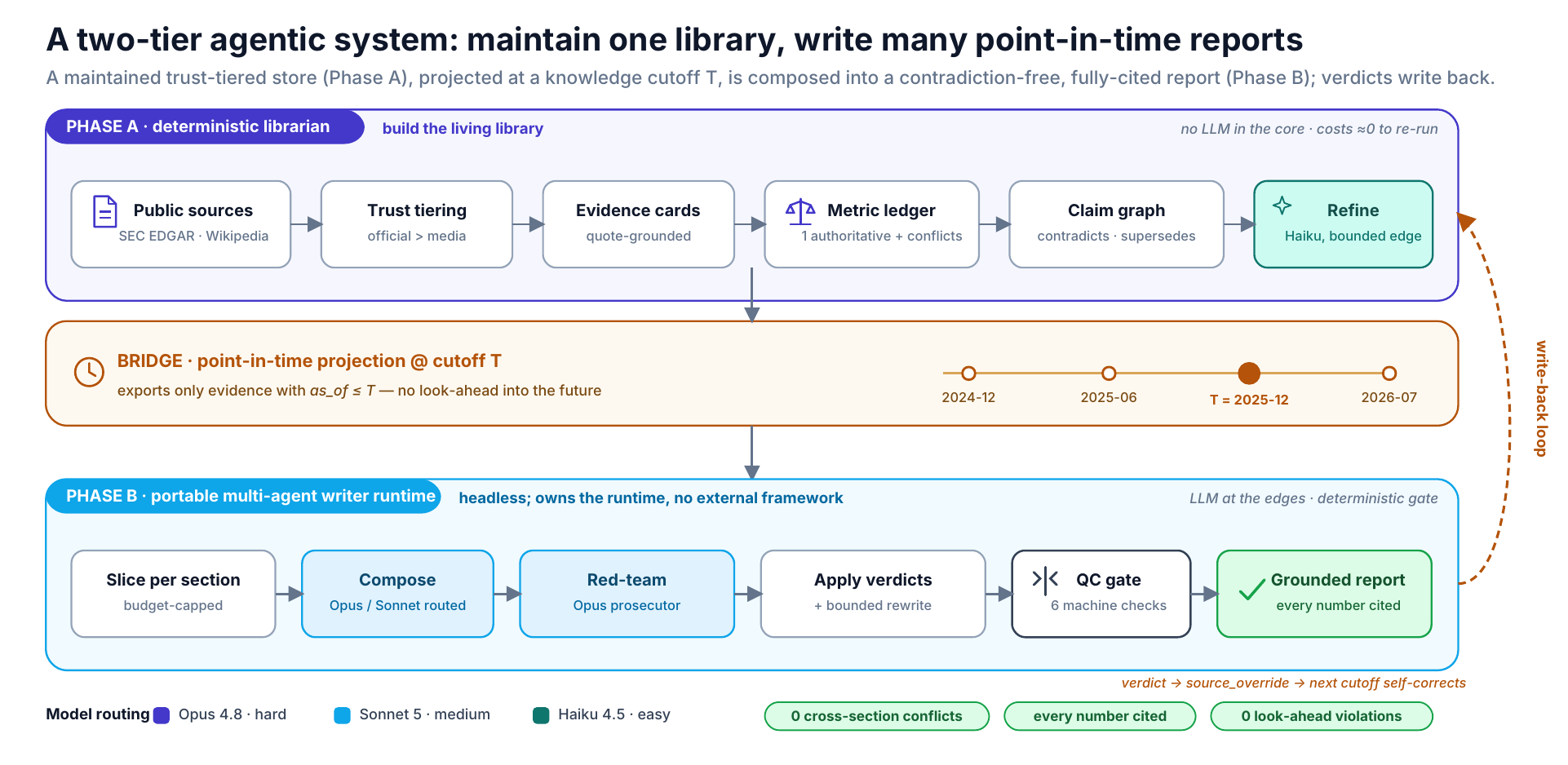}
  \caption{End-to-end system. \textbf{Phase~A} (deterministic librarian) ingests
  timestamped public sources into trust-tiered evidence cards, a metric ledger, and
  a claim graph. The \textbf{cutoff dial} selects a knowledge time $T$; only
  evidence with $\mathit{as\_of}\le T$ is exported. \textbf{Phase~B} (portable
  agent runtime) slices per section, composes with difficulty-tiered model routing,
  red-teams, and passes a deterministic QC gate. Red-team verdicts feed back to the
  store.}
  \label{fig:e2e}
\end{figure*}

The system runs as two decoupled phases coupled by a timestamped store and closed
by a write-back loop (Fig.~\ref{fig:e2e}).

\paragraph{Phase A: the librarian (deterministic).}
A pipeline of deterministic Python stages ingests public sources, recording each
one's true publication date and trust tier. The librarian maintains knowledge at
three progressively-distilled layers: quote-grounded \emph{evidence
cards} (raw facts, each pinned to a source quote), an \emph{authoritative metric
ledger} (one reconciled value per company-metric pair), and a \emph{claim graph} of
\texttt{contradicts}, \texttt{supersedes}, and \texttt{qualifies} edges over those
values. Each layer is derived deterministically from the one below, so a number
traces down to its source quote and up to its conflicts. The core costs essentially
nothing to re-run; an LLM is used only at one clearly-bounded
seam: a cheap refinement pass (Claude Haiku~4.5) that corrects a numeric card's
value/unit against its own quote or demotes it to qualitative. The headline metrics,
computed deterministically, are thus reproducible given a fixed store.
Appendix~\ref{app:artifact} traces one real metric through all three layers and into
a delivered report; Appendix~\ref{app:schema} gives their required fields.

\paragraph{The bridge: point-in-time projection.}
Given a cutoff~$T$, the bridge projects the store into the writer's four artifacts
(outline, evidence cards, metric ledger, claim graph), filtering to
$\mathit{as\_of}\le T$. This is the no-look-ahead seam: a report ``as of'' a past
date sees exactly the evidence that existed then. Object mapping is near 1:1 (both
tiers speak the same three layers), so the bridge is a data adapter, not a rewrite.

\paragraph{Phase B: the portable writer runtime.}
The writer is a self-contained, headless runtime over a tiered LLM provider and a
bounded-con\-currency pool, with no dependency on an external agent framework, so it
embeds equally in a service or a batch job. Its orchestration wraps a set of
deterministic scripts (slice, tag-normalize, QC, render) around the LLM calls. The
workflow is a fixed directed acyclic graph (DAG):
slice each section~$\to$ compose (one LLM call per section, tier-routed)~$\to$
normalize~$\to$ red-team (a Claude Opus~4.8 ``prosecutor'' per section that returns
holds/weak/refuted verdicts)~$\to$ apply verdicts~$\to$ a bounded rewrite of
affected sections~$\to$ deterministic convergence backstop~$\to$ QC gate~$\to$
render. We use standard multi-agent building blocks (file-per-agent artifacts,
contract-first outlines, tool-mediated deterministic checks) but implement them in a
runtime of our own, without binding to any external framework~\cite{anthropic2024multiagent}.

\paragraph{Write-back loop.}
When the red-team refutes a card, the verdict maps to a librarian
\texttt{source\_override}/claim \texttt{refuted}, so the next run at the next cutoff
inherits the correction: the store is a living library, not a static dump. The
write-back is guarded: an override never invents a value, it only demotes the refuted
source so the ledger falls back to a pre-existing same-kind alternative, and every
change appends a row to an append-only audit log, so promotion is
human-gated and reversible. Regeneration is idempotent: it carries a human promotion
forward rather than resetting it, but snapshots the exact evidence approved, so if
that evidence later vanishes, even when back-filled cards keep the count unchanged (an
``anchor swap''), the claim is flagged for re-validation, not silently kept.
Algorithm~\ref{alg:e2e} states the two phases end to end.

\subsection{The distributed multi-agent design}
\label{sec:distributed}
Phase~B is where the system is genuinely \emph{distributed}, by deliberate design.
\textbf{(1) Parallel per-section agents.} The report outline is a
partition: each section is an independent compose task, and the sections fan out
across a bounded-concurrency pool (\S\ref{sec:trust}) rather than being written
serially. Sections are the natural unit of parallelism because the outline contract
makes them near-independent: the only shared state is the metric ledger, read-only
at compose time. \textbf{(2) Heterogeneous agents.} A difficulty router sends
conflict-touching sections to a stronger, costlier model (Opus) and routine sections
to a cheaper one (Sonnet), so compute is spent where the reasoning is hard: classic
heterogeneous scheduling, applied to LLM agents. \textbf{(3) Separation of powers.}
Composition and criticism are \emph{different} agents with opposing objectives: a
composer writes to satisfy the section contract, an independent Opus ``prosecutor''
red-teams the draft to break it. Neither grades its own work, and the arbiter that
decides delivery is the deterministic QC gate, not an LLM. \textbf{(4) Coordination
through a shared store, not messages.} The agents never talk to each other directly;
they coordinate \emph{stigmergically} through the trust-tiered store: composers read
the same authoritative ledger, and the red-team writes verdicts back to it. This
adds parallelism without the usual multi-agent failure mode of concurrent writers
diverging: because the single authoritative value lives in the shared ledger rather
than in each agent's context, two sections physically cannot cite different numbers
for the same metric. The distributed design thus buys throughput
(\S\ref{sec:eval}, E6) \emph{without} paying in consistency. Figure~\ref{fig:distributed}
(App.~\ref{app:distributed}) draws the four mechanisms as one picture.

\emph{Why the fan-out is safe.} The consistency guarantee is structural, not a
matter of scheduling. The bridge (\S\ref{sec:system}) emits an \emph{immutable,
point-in-time snapshot} at cutoff~$T$, and every composer reads from that single
snapshot, so within one run there are no concurrent writers: no lock, barrier, or
two-phase commit is needed, and the classic shared-memory races (write--write, torn
reads, deadlock) cannot arise. The only writer is the red-team's write-back, deferred
to the \emph{next} cutoff, never mid-run. Read-only fan-out over an immutable
snapshot makes each compose step idempotent and order-independent, so the worker
count~$K$ trades latency against cost (E6) but cannot change the delivered numbers.

\begin{algorithm}[t]
\small
\DontPrintSemicolon
\SetKwInOut{Input}{input}\SetKwInOut{Output}{output}
\SetKwFunction{Slice}{Slice}\SetKwFunction{Compose}{Compose}
\SetKwFunction{RedTeam}{RedTeam}\SetKwFunction{QC}{QC}\SetKwFunction{WriteBack}{WriteBack}
\Input{public sources $\mathcal{S}$; cutoff $T$; outline $\mathcal{O}$}
\Output{grounded report $R$; updated store $\mathcal{L}$}
\tcp{\textbf{Phase A}: deterministic librarian (LLM-free core)}
\ForEach{$s \in \mathcal{S}$}{
  tier, $\mathit{as\_of} \leftarrow$ classify$(s)$\;
  cards $\leftarrow$ extract quote-grounded evidence from $s$\;
  \lIf{numeric \textnormal{\&} ambiguous}{Haiku-refine card vs.\ its quote}
}
$\mathrm{ledger} \leftarrow$ per (co., metric): pick by tier $\to$ corrob.\ $\to$ recency\;
$\mathrm{graph} \leftarrow$ contradicts / supersedes / qualifies edges\;
\tcp{\textbf{Bridge}: point-in-time projection (no look-ahead)}
$E_T \leftarrow \{\, c \in \mathcal{L} : \mathit{as\_of}(c) \le T \,\}$; recompute ledger @ $T$\;
\tcp{\textbf{Phase B}: portable multi-agent writer}
\ForEach{section $\sigma \in \mathcal{O}$ \textnormal{(bounded parallel)}}{
  $C_\sigma \leftarrow$ \Slice{$E_T,\sigma$} capped by salience budget\;
  $d_\sigma \leftarrow$ Opus if $\sigma$ touches unresolved conflict else Sonnet\;
  $\mathrm{draft}_\sigma \leftarrow$ normalize(\Compose{$C_\sigma$; model $d_\sigma$})\;
  $v_\sigma \leftarrow$ \RedTeam{$\mathrm{draft}_\sigma$} \tcp*{Opus prosecutor}
}
apply verdicts; bounded rewrite; deterministic convergence backstop\;
\While{\QC{$R$} $\neq \emptyset$ \textnormal{and} rounds $<$ cap}{rewrite flagged sections}
\lForEach{refuted $v_\sigma$}{\WriteBack{$\mathcal{L}$} \tcp*{\footnotesize source\_override}}
\Return render($R$), $\mathcal{L}$\;
\caption{Maintain library, write report @ $T$.}
\label{alg:e2e}
\end{algorithm}

\section{Trust \& Consistency Mechanisms}
\label{sec:trust}
\paragraph{Source tiers and permitted use.}
\begin{sloppypar}
Every source is typed from text/path cues and assigned a trust tier together with a
\emph{permitted use} that governs whether its numbers may be cited: U.S. Securities
and Exchange Commission (SEC) filings become \emph{official} (usable as hard
evidence), U.S. Bureau of Labor Statistics (BLS) macro-statistics releases become
\emph{gov\_stat} (supporting evidence: authoritative, but for macro context, not a
company's own figures), and Wikipedia becomes \emph{media} (routing only). The tiers
are strictly ordered (official~$>$~gov\_stat~$>$~sell\_side~$>$~media; \emph{sell\_side}
is analyst research, used in company deployments but omitted here as it is not
redistributable, so it appears only in E4's gold set): routing-only
sources may inform entity/topic routing and context but can never become a citable
hard-evidence number, and a supporting-tier macro value can never displace a
company's own official figure: the ``official-first'' rule.
\end{sloppypar}

\paragraph{Metric ledger.}
For each (company, metric) the ledger selects one authoritative value by a fixed
policy: \emph{tier} dominates, then \emph{corroboration} (distinct-source count)
breaks ties within a tier, then \emph{recency} ($\mathit{as\_of}$). Competing
values are retained as alternatives and flagged as a conflict when a comparable
same-kind figure materially disagrees (a fixed $>\!15\%$ threshold, so unit-equal
restatements do not spuriously fire), but the report cites the single authoritative
value, which is what eliminates cross-section drift.

\paragraph{QC gate (six deterministic checks).}
The delivery gate is language-neutral and LLM-free: (1) orphan citations, (2)
unsourced numbers, (3) numeric drift across sections, (4) buried
contradictions (a claim-graph conflict whose two endpoints are not reconciled
together), (5) unregistered metrics, (6) cross-section contradiction. A report is
deliverable only when the error set is empty (the six checks are specified in
Appendix~\ref{app:qc}). Because this gate computes the
headline metrics, we validate the gate itself (\S\ref{sec:eval}, E5).

\section{Deployment \& Dataset}
\label{sec:dataset}
We self-collected a public, redistributable, English-only corpus at
\emph{production scale} (Table~\ref{tab:corpus}): \textbf{6{,}130 sources}
extracting to \textbf{555{,}926 evidence cards} (457{,}561 numeric) and a metric
ledger of 2{,}589 authoritative company-metric values, 2{,}132 of them carrying
recorded conflicts. The three tiers are 5{,}397 SEC EDGAR filings~\cite{sec_edgar}
(official) for 295 issuers across 11 sectors, 672 U.S. Bureau of Labor Statistics
(BLS) macro releases (gov\_stat: authoritative supporting context that, carrying no
company attribution, by design never enters the per-company ledger), and 61 Wikipedia
articles (media: routing and context only, by design backing no numbers). Filings are
fetched keyless via the official EDGAR and BLS REST APIs with provenance headers.

The design point is \textbf{one library, many reports}: a single maintained store
serves multiple report theses rather than being purpose-built for one. From this
store we generate four flagship point-in-time reports at a common cutoff
(2025-12-31) (AI-compute, energy, healthcare/pharma, and banks), each projected by
the sector-scoped bridge, and each passing the QC gate with zero errors. Breadth
matters for the evaluation too: the corpus contains thousands of
naturally-occurring cross-period and cross-source contradictions (drift in remaining
performance obligations, backlog, and capital expenditure; restatements; revised
guidance) across every sector, never fabricated drift. Each source is retained under its issuing API's terms, with per-tier
licensing recorded.

\begin{table}[t]
  \centering
  \caption{Self-collected corpus (public, timestamped, multi-tier, multi-sector).}
  \label{tab:corpus}
  \small
  
\begin{tabular}{lr}
\toprule
Property & Value \\
\midrule
Sources registered (total) & 6130 \\
\quad official (SEC filings, hard evidence) & 5397 \\
\quad gov\_stat (BLS macro, supporting) & 672 \\
\quad media (Wikipedia, routing only) & 61 \\
\quad of which produced $\ge$1 card & 6054 \\
Sectors covered & 11 \\
Companies (issuers) & 295 \\
Evidence cards & 555926 \\
Numeric cards & 457561 \\
Media$\to$numeric leakage & 0 \\
Publication span & 2023-06 -- 2026-07 \\
\bottomrule
\end{tabular}

\end{table}

Because every source carries its true publication date, we replay time by filtering
the store to a cutoff (Phase~B), mimicking ``30/60/90 days later'' without
re-fetching: the production case where later filings revise or contradict a value an
earlier report relied on.

\section{Evaluation}
\label{sec:eval}
Every headline metric is machine-computed (no LLM decides a reported number) and,
where an ablation applies, compared on the \emph{identical} corpus against an
explicit baseline arm. Deterministic experiments (E1, E3, E4, E5, E7, E8) are exact
by construction; the LLM-dependent cost/latency/quality comparison (E6) is run on
the real Bedrock backend over three repeats. All numbers in the tables are emitted
directly from the stored per-experiment result records (reproduction steps in
Appendix~\ref{app:repro}); the headline results are also charted together in
Fig.~\ref{fig:results} (App.~\ref{app:results}).

\paragraph{E1: a shared ledger removes cross-section drift.}
Without a shared ledger, a writer grounding each section independently surfaces
every competing value for a metric; the ledger collapses each to one authoritative
value. On the real store at the final cutoff, the no-ledger baseline would emit
\textbf{6{,}845} contradictory figures across 2{,}105 metrics with competing
values; ours emits \textbf{0} (Table~\ref{tab:e1}). Replayed across seven cutoffs, the ledger's
authoritative value changed \textbf{4{,}732} times, and \emph{every} change was
justified by newer, higher-tier, or more-corroborated evidence (0 unexplained).

\paragraph{E2: grounding.}
Across all four flagship theses composed from the one library (AI-compute, energy,
healthcare/pharma, banks; cutoff 2025-12-31), every numeric-bearing body line must
carry an evidence citation or a metric annotation. Aggregate grounding is
\textbf{202/203} (\textbf{99.5\%}) with \textbf{0} orphan citations and \textbf{0}
unregistered metrics; three reports are 100\%, and the lone exception is a synthesis
sentence whose figures are each cited earlier in the same section.

\paragraph{E3: trust tiering suppresses rumor and quarantines macro context.}
End-to-end on the real store, all three tiers classify correctly (5{,}397 filings
as \texttt{hard\_evidence}, 672 BLS releases as \texttt{supporting\_\allowbreak evidence}
(gov\_stat), 61 Wikipedia articles as \texttt{routing\_only}, \textbf{0}
misclassified), and \textbf{0} of the 457{,}561 numeric cards trace to a
routing-only source. The third tier is genuinely mined (2{,}352 numeric gov\_stat
cards: CPI, PPI, payrolls, unemployment), yet because a macro statistic carries no
company attribution, \textbf{0} gov\_stat values displace a company's own official
figure and the per-company ledger stays 100\% official: neither media nor macro
statistics ever become a citable number, and the invariant holds across the full
production-scale corpus (all 457{,}561 numeric cards, 295 issuers, 11 sectors).

\paragraph{E4: tier-first selection beats popularity.}
We run two selection policies on one labeled gold set of metric clusters: our
tier-first ledger, and a popularity-first baseline that takes the value with the
most distinct backing sources (the ``most-cited''/semantic-layer heuristic),
ignoring tier. Tier-first is correct on \textbf{22/22} cases; popularity-first
scores only \textbf{9/22} (Table~\ref{tab:e4}). Thirteen cases are
\emph{popularity traps}: a widely-repeated lower-tier value (a rumor echoed by
several media sources, or a corroborated macro statistic) competes with a single
official filing; the tier rule survives all thirteen, popularity adopts the wrong
value every time. The set spans the full configured lattice
(official~$>$~gov\_stat~$>$~sell\_side~$>$~media), including the invariant that a
newer, more-corroborated gov\_stat value still cannot displace an official figure
(which gov\_stat may anchor only when none exists), and corroboration breaking ties
only \emph{within} a tier. It is a \emph{designed coverage lattice}, not a sample:
the cross-tier conflict it probes cannot arise on this corpus (all \textbf{2{,}589}
real ledger clusters are single-tier, as gov\_stat cards carry no issuer name and
media is never mined for a number), yet the within-tier rule is exercised at scale:
over \textbf{2{,}132} real conflicts our corroboration$\rightarrow$recency policy
differs from naive newest-wins on \textbf{973}, far beyond the 22 gold cases.

\paragraph{E5: the checker is trustworthy (recall \emph{and} precision).}
A gate is only trustworthy if it both catches real defects and stays quiet on clean
text. We clone a clean, QC-passing run and (i) inject five defect classes (orphan
citation, unsourced number, broken cross-reference, unregistered metric, buried
contradiction) and (ii) apply three \emph{negative controls}: defect-free
perturbations that must \emph{not} fire (a paragraph reusing only already-valid
citations and metric tokens, a duplicated grounded line, number-free prose). Recall
is \textbf{1.0} (5/5) and precision is \textbf{1.0} with a \textbf{0} false-positive
rate on the controls (Table~\ref{tab:e5}). Crucially, we separate delivery-blocking
from advisory detection: the delivery gate is ``error set empty,'' and \textbf{3/3}
error-level defects raise a blocking error, while the two warning-level defects are
caught but advisory by design. This disarms ``graded your own homework'': the gate
that computes E1--E4 is itself validated on both axes.

\paragraph{E6: in the distributed writer, parallelism and routing cut cost at comparable quality.}
Isolating the multi-agent compose fan-out on identical slices over three repeats
(Table~\ref{tab:e6}), bounded-concurrency parallelism across the per-section agents
runs \textbf{$3.7\times$} faster than serial, and difficulty-tiered routing
(conflict-touching sections~$\to$~Opus, the rest~$\to$~Claude Sonnet~5) costs
\textbf{4.1\% less} than sending every section to Opus.
The cost gap is deliberately modest on this flagship: it is conflict-heavy, so
5~of~6 sections touch an unresolved edge and correctly route to Opus, so routing
saves little precisely when the report is hard, and the same dial saves far more on a
low-conflict thesis. All four variants pass
QC with zero errors, but a binary gate cannot rank them, and all-Sonnet is the
cheapest, so ``equal quality'' needs an independent signal. We add a deterministic,
graded quality score (grounding coverage $+$ conflict-pair coverage $+$
output-contract adherence, computed on each variant's actual prose, independent of
the QC gate). The counterintuitive result: tiered routing scores \textbf{above} the
all-Opus ceiling ($+0.079$) and far above all-Sonnet ($+0.262$). Spending the strong
model only where reasoning is hard beats spending it everywhere, so the cheaper
all-medium point is not the default: it saves on easy sections but degrades exactly
the conflict-heavy synthesis that routes to Opus.

\begin{figure}[t]
  \centering
  \includegraphics[width=0.80\columnwidth]{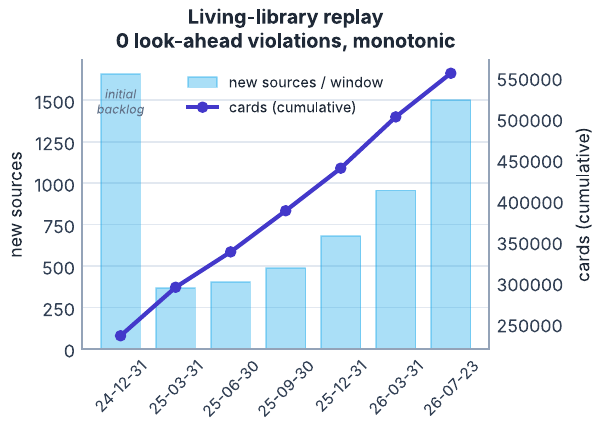}
  \caption{E7: point-in-time replay across seven cutoffs. Bars (left axis) count
  sources newly available in each window; the line (right axis) is the cumulative
  evidence-card base, which grows monotonically with zero look-ahead (a report as
  of $T$ reads only \mbox{$\mathit{as\_of}\le T$}). The first bar is the pre-existing
  back-catalog loaded at the opening cutoff, not a per-window rate.}
  \label{fig:e7}
\end{figure}

\paragraph{E7: the living library grows without look-ahead.}
Replaying the store across seven cutoffs (Table~\ref{tab:e7},
Fig.~\ref{fig:e7}) yields
\textbf{0} look-ahead violations and monotonic growth
(235{,}373$\to$555{,}312 cards; 1{,}659$\to$6{,}054 card-bearing sources), capturing
\textbf{4{,}395} post-initial evidence-arrival events (new filings, restatements,
new conflicts) and lifting recorded metric conflicts from 1{,}770 to 2{,}132. This closes
end to end at the report level: regenerating the AI-compute flagship at three
advancing cutoffs yields a report that grows in lockstep (27{,}104$\to$42{,}900
cards, 255$\to$276 metrics, 562$\to$655 reconciled conflict edges) while every cutoff
stays deliverable (QC errors~$=0$), and all four flagship theses compose at the
shared 2025-12-31 cutoff from this one library. A static one-shot corpus cannot
exhibit this: as real evidence arrives, the library grows and corroboration rises.

\paragraph{E8: the write-back loop self-corrects a later run.}
Growth is only half of ``living''; the loop must also close on \emph{corrections}.
We trace one worked case end to end using the librarian's real override machinery
(Table~\ref{tab:e8}, App.~\ref{app:tables}). In this illustrative scenario the
report-side red-team challenges the authoritative interest figure (\$19mn) after
flagging its backing filing as low-confidence. The verdict maps to a librarian \texttt{source\_override}
(status \texttt{retracted}); on re-ingestion the 284 evidence cards from that filing
inherit the non-active status, and because the metric ledger considers only
active-source cards, the authoritative value self-corrects to \$9mn, an already-recorded
alternative, with \textbf{0} manual value edits. Tracing the same override path on a
batch of auto-discovered conflicted metrics, 5 of 6 write-backs self-correct, each to
a pre-existing same-kind alternative (0 manual edits), so the loop closes on many
values, not one, deterministically and auditably.

\begin{table}[t]
  \centering
  \caption{E1: cross-section figure drift (lower is better).}
  \label{tab:e1}
  \small
  
\begin{tabular}{lrrr}
\toprule
Condition & \shortstack{drift\\figs} & \shortstack{metrics w/\\conflict} & \shortstack{justified\\drift} \\
\midrule
Baseline (no shared ledger) & 6845 & 2105 & --- \\
\textbf{Ours (shared ledger)} & \textbf{0} & 2105 & --- \\
\midrule
\multicolumn{4}{l}{\footnotesize Temporal drift, 7 cutoffs: 4732 changes, 100\% evidence-justified, 0 unexplained.} \\
\bottomrule
\end{tabular}

\end{table}

\begin{table}[t]
  \centering
  \caption{E4: authoritative-value selection vs.\ gold.}
  \label{tab:e4}
  \small
  
\setlength{\tabcolsep}{4pt}
\begin{tabular}{lcc}
\toprule
Gold case & ours & popularity-first \\
\midrule
official beats 3$\times$ media rumor$^{t}$ & \checkmark & \ding{55} \\
official beats newer 2x sell$^\dagger$$^{t}$ & \checkmark & \ding{55} \\
newer official wins tie & \checkmark & \checkmark \\
corrob tie-break & \checkmark & \checkmark \\
sell beats 2x media$^\dagger$$^{t}$ & \checkmark & \ding{55} \\
corrob $>$ newer single & \checkmark & \checkmark \\
gov beats 3$\times$ media rumor$^{t}$ & \checkmark & \ding{55} \\
official beats newer gov$^{t}$ & \checkmark & \ding{55} \\
official beats 2$\times$ gov$^{t}$ & \checkmark & \ding{55} \\
gov fills absent official & \checkmark & \checkmark \\
gov corrob tie-break & \checkmark & \checkmark \\
newer gov revision wins & \checkmark & \checkmark \\
official beats 4$\times$ media rumor$^{t}$ & \checkmark & \ding{55} \\
official beats 3$\times$ sell$^\dagger$$^{t}$ & \checkmark & \ding{55} \\
gov beats 2$\times$ media$^{t}$ & \checkmark & \ding{55} \\
modal guards \$-misparse & \checkmark & \checkmark \\
gov beats 2$\times$ sell$^\dagger$$^{t}$ & \checkmark & \ding{55} \\
official $>$ gov $>$ media (lattice)$^{t}$ & \checkmark & \ding{55} \\
newer official on equal corrob & \checkmark & \checkmark \\
kind guard drops stray percent & \checkmark & \checkmark \\
official beats 5$\times$ media crowd$^{t}$ & \checkmark & \ding{55} \\
official beats corrob newer gov$^{t}$ & \checkmark & \ding{55} \\
\midrule
\textbf{Accuracy} & \textbf{22/22} & 9/22 \\
\multicolumn{3}{@{}p{\columnwidth}@{}}{\footnotesize ours $=$ tier-first (ledger policy). $^{t}$13 popularity traps (widely-repeated lower-tier value); tier rule survives all 13. $^\dagger$configured-but-unused tier.} \\
\bottomrule
\end{tabular}

\end{table}

\begin{table}[t]
  \centering
  \caption{E5: QC gate meta-evaluation, recall and precision (defect injection
  plus negative controls).}
  \label{tab:e5}
  \small
  \renewcommand{\arraystretch}{0.92}
  
\begin{tabular}{lll}
\toprule
Perturbation & level & QC outcome \\
\midrule
orphan citation & error & caught, blocks delivery \\
unsourced number & warning & caught, advisory (warning) \\
broken xref & warning & caught, advisory (warning) \\
unregistered metric & error & caught, blocks delivery \\
buried contradiction & error & caught, blocks delivery \\
\midrule
\;neg: reuse existing citations & (clean) & 0 new errors \\
\;neg: duplicate grounded line & (clean) & 0 new errors \\
\;neg: prose only no numbers & (clean) & 0 new errors \\
\midrule
\multicolumn{3}{l}{\textbf{Recall 5/5=1.00}, \textbf{precision 1.00}, FP-rate 0.00} \\
\multicolumn{3}{l}{\footnotesize over 3 controls; 3/3 error-level defects block delivery (rest advisory).} \\
\bottomrule
\end{tabular}

\end{table}

\begin{table}[t]
  \centering
  \caption{E6: cost/latency of routing and parallelism (compose fan-out, real
  Bedrock).}
  \label{tab:e6}
  \small
  \renewcommand{\arraystretch}{0.92}
  
\begin{tabular}{llrrr}
\toprule
Configuration & mix & \$/rep & wall (s) & quality \\
\midrule
All-hard (Opus), par. & 6H/0M & \$2.35 & 64 & 0.714 \\
All-medium (Sonnet), par. & 0H/6M & \$0.49 & 47 & 0.531 \\
Tiered, serial & 5H/1M & \$2.26 & 184 & 0.801 \\
\textbf{Tiered, par.\ (ours)} & 5H/1M & \textbf{\$2.25} & \textbf{50} & \textbf{0.793} \\
\midrule
\multicolumn{5}{@{}p{\columnwidth}@{}}{\footnotesize Mean over 3 repeats. Tiered vs all-hard: 4.1\% cost saved, 3.7$\times$ faster than serial. Quality (grounding+conflict+contract, indep.\ of QC): tiered$-$all-hard $=+0.079$, $-$all-medium $=+0.262$; all pass QC.} \\
\bottomrule
\end{tabular}

\end{table}

\begin{table}[t]
  \centering
  \caption{E7: living-library replay across cutoffs.}
  \label{tab:e7}
  \small
  \renewcommand{\arraystretch}{0.92}
  
\begin{tabular}{lrrrrr}
\toprule
Cutoff $T$ & cards & sources & metrics & conflicts & new src \\
\midrule
2024-12-31 & 235373 & 1659 & 2279 & 1770 & 1659 \\
2025-03-31 & 294477 & 2026 & 2343 & 1868 & 367 \\
2025-06-30 & 337597 & 2430 & 2388 & 1921 & 404 \\
2025-09-30 & 387741 & 2916 & 2462 & 2001 & 486 \\
2025-12-31 & 439605 & 3597 & 2493 & 2026 & 681 \\
2026-03-31 & 502134 & 4553 & 2544 & 2093 & 956 \\
2026-07-23 & 555312 & 6054 & 2589 & 2132 & 1501 \\
\midrule
\multicolumn{6}{l}{\footnotesize 0 look-ahead violations; monotonic. New-src sums to 6054 card-bearing sources;} \\
\multicolumn{6}{l}{\footnotesize 4395 post-initial arrivals (excl.\ the 1659 at the first cutoff).} \\
\bottomrule
\end{tabular}

\end{table}

\paragraph{Limitations.}
The corpus is English-only and three-tier as collected (official SEC filings,
gov\_stat BLS macro statistics, media Wikipedia; sell\_side omitted, \S\ref{sec:trust}).
The gov\_stat tier is authoritative but macro-only: its releases carry no
company attribution, so by construction it enriches context, not the per-company
ledger. Entity linking is substring-based, so it occasionally over-attributes a
metric when a company's short name is a substring of unrelated filing text (e.g.\
``3M''); this is an auditable extraction artifact, not a ledger-policy error. The E4
cross-tier gold set is designed rather than sampled (the deployed corpus has no
cross-tier clusters), though the within-tier rule is corroborated on 2{,}132 real
conflicts; E6 uses a deterministic quality proxy, not human judgement; and E8 traces
write-back on a small batch. E1's no-ledger arm isolates the ledger's effect, not a
strong shared-state competitor (a graph- or semantic-layer-backed retriever), the
natural next comparison our design is built to host.

\section{Lessons \& Related Work}
\label{sec:lessons}
\paragraph{Lessons.}
(i) \emph{Govern by trust, not popularity}: a most-cited-value heuristic adopts
widely-repeated rumor, while an official-first ledger is simpler and correct
(E1, E3, E4). (ii) \emph{Keep the core deterministic, put the LLM at the edges}:
deterministic selection, consistency, and QC make the headline metrics reproducible
and the gate meta-evaluable (E5). (iii) \emph{Own the runtime}: standard multi-agent
building blocks (file-per-agent artifacts, contract-first outlines) carry over cleanly to
a headless service without binding to any external agent framework. (iv) \emph{Routing buys quality, not just cheapness}: tiered
composition matches or beats the all-Opus ceiling while a uniform-cheap baseline
degrades, at modest dollar saving on a conflict-heavy report (E6).

\paragraph{Related work.}
\emph{Grounding and citation.}
RAG grounds individual answers~\cite{lewis2020rag,gao2023ragsurvey} per query rather
than maintaining a reconciled ontology, enforcing no cross-time consistency;
Self-RAG~\cite{asai2024selfrag} adds a self-critique, the same model judging its own
output, not a deterministic, meta-validated gate.
Grounding metrics like FActScore~\cite{min2023factscore} and RAGAs~\cite{es2023ragas},
citation benchmarks like ALCE~\cite{gao2023alce}, and rubrics like
FinReasoning~\cite{zhu2026finreasoning} \emph{score} whether a text is supported but
do not \emph{govern} a living store; our QC gate blocks delivery, not a post-hoc
score. STORM~\cite{shao2024storm} synthesizes Wikipedia-like articles from retrieved
sources but has no source-authority tier, metric ledger, or point-in-time discipline.

\emph{Temporal and streaming settings.}
Closest is \emph{temporal/streaming QA}: TimeQA~\cite{chen2021timeqa},
StreamingQA~\cite{liska2022streamingqa}, and RealTime~QA~\cite{kasai2023realtimeqa}
isolate look-ahead and evolving knowledge, but for \emph{short-answer} questions; we
target long-form reports whose challenge is \emph{cross-section} consistency.
GraphRAG~\cite{edge2024graphrag} and GFM-RAG~\cite{luo2025gfmrag} reason over an
entity/claim graph but carry no provenance-tier or point-in-time projection,
reconciling per query not in a maintained store; MemGPT~\cite{packer2023memgpt}
persists agent state without source-authority or no-look-ahead discipline.

\emph{Agentic frameworks and verification.}
Agentic frameworks \cite{wu2023autogen,hong2023metagpt,yao2023react,shinn2023reflexion}
and toolkits such as LangGraph~\cite{langgraph2024} provide the stateful multi-actor
graph we deliberately \emph{do not} bind to, building a minimal headless
orchestrator of our own instead (Lesson~iii). Verified orchestration~\cite{zhang2026verified}
closes a plan-verify-replan loop like ours, but we keep the trust backbone and
delivery gate deterministic and meta-validated (E5), with human-gated promotion via
an append-only log (E8).
Registry-driven grounding like REGAL~\cite{agrawal2026regal} and ontology-guided
extraction like OntoMetric~\cite{yu2025ontometric} share our deterministic-core
discipline but ground structured telemetry or a single-document ESG graph, not
\emph{unstructured, cross-document} evidence reconciled into a timestamped ledger.
Financial QA benchmarks~\cite{chen2021finqa,islam2023financebench} and
FinCARDS~\cite{zhou2026fincards} target intra-document reasoning; semantic-layer
analytics~\cite{databricks2024genie} share the ``governed source of truth'' intuition
we extend to timestamped, trust-tiered evidence, with
LLM-as-judge~\cite{zheng2023judge} used only as a red-team.

\section{Conclusion}
Separating a maintained, trust-tiered, point-in-time library from report writing
turns long-form generation's chronic drift, provenance loss, and trust flattening
into mechanically-checkable, largely eliminated properties.

\balance

\clearpage
\appendix

\section{Worked artifact: one metric, end to end}
\label{app:artifact}
We trace a single real metric, Oracle's Remaining Performance Obligations
(RPO), through the three librarian artifacts and into the delivered report, to
make the data model concrete. All snippets are verbatim from the store and a
delivered run (identifiers abbreviated for space).

\paragraph{(A) Evidence card.} The deterministic extractor emits one quote-grounded
card per (source, metric) hit. The card carries the verbatim quote, so every
downstream number is auditable back to the filing text:
\begin{lstlisting}[language=,caption={Evidence card (SEC 10-K, official tier).},captionpos=b]
{ "evidence_id": "ev_orcl_..._rpo_d0c7b8b4",
  "source_id":   "src_orcl_..._10_k",
  "company": "Oracle",  "metric": "RPO",
  "quote": "Remaining performance obligations were
     $638 billion and $138 billion as of May 31,
     2026 and 2025, respectively.",
  "metric_value": "$638 billion",  "value_norm": 638000.0,
  "value_kind": "money_mn",  "source_tier": "official",
  "as_of": "2026-06-22",  "evidence_kind": "quantitative",
  "source_status": "active" }
\end{lstlisting}

\paragraph{(B) Metric ledger.} For each (company, metric) the ledger selects one
authoritative value by tier~$\to$~corroboration~$\to$~recency and retains the
losers as \texttt{alternatives} with a \texttt{disagrees} flag: this is what a
section cites, and what makes cross-section drift impossible:
\begin{lstlisting}[language=,caption={Metric-ledger row with retained alternatives.},captionpos=b]
{ "metric_id": "mtr_oracle_rpo",  "company": "Oracle",
  "metric": "RPO",  "authoritative_value": "$638.0bn",
  "value_norm": 638000.0,  "as_of": "2026-06-22",
  "source_tier": "official",  "value_conflict": true,
  "basis_evidence_id": "ev_orcl_..._rpo_d0c7b8b4",
  "alternatives": [
    {"value": "$552.6bn", "as_of": "2026-03-11", "disagrees": false},
    {"value": "$523.3bn", "as_of": "2025-12-11", "disagrees": true},
    {"value": "$455.3bn", "as_of": "2025-09-10", "disagrees": true} ] }
\end{lstlisting}

\paragraph{(C) Delivered report prose.} The composed section writes numbers not as
literals but as symbolic ledger handles (e.g.\ \texttt{\{m\_oracle\_\allowbreak rpo:\allowbreak authoritative\}})
that are substituted at render, and cites evidence-card ids inline (\texttt{[E14847]}),
so no number is typed by hand and every one is traceable. The rendered text reads:
\begin{quote}\small\itshape
``As of May 31, 2026, Oracle reported remaining performance obligations (RPO)
of \$638.0bn, up sharply from \$455.3bn three quarters
earlier [E14847][E14761]. \ldots{} Oracle expects to recognize only
about 10\% over the next twelve months [E14847].''
\end{quote}

\paragraph{(D) Claim-graph edge.} Cross-source and cross-period tensions are stored
as typed edges; a \texttt{supersedes} edge is what lets a later filing override an
earlier value while both stay auditable:
\begin{lstlisting}[language=,caption={Claim-graph edge (supersedes).},captionpos=b]
{ "edge": "supersedes",  "metric": "RPO",  "company": "Oracle",
  "from_evidence": "ev_orcl_..._rpo_d0c7b8b4",   // $638.0bn @ 2026-06-22
  "to_evidence":   "ev_orcl_..._rpo_e1366a04",   // $455.3bn @ 2025-09-10
  "reason": "newer official filing, same metric" }
\end{lstlisting}

\section{The QC gate (six deterministic checks)}
\label{app:qc}
The delivery gate is language-neutral and LLM-free; a report is deliverable only
when the error set is empty. Checks (1) and (4)--(6) raise blocking errors; (2) is
advisory (warning-level). Check (3) cannot fire once one authoritative value backs
every annotation (the drift E1 removes structurally). E5 validates the gate by
defect injection.
\begin{enumerate}[leftmargin=1.4em,itemsep=1pt]
\item \textbf{Orphan citation}: a citation marker with no backing evidence card.
\item \textbf{Unsourced number}: a numeric body line with neither an evidence
citation nor a metric annotation.
\item \textbf{Numeric drift}: the same metric rendered with two different
values across sections (the drift E1 eliminates).
\item \textbf{Buried contradiction}: a claim-graph conflict whose two endpoints
are not reconciled in the same place.
\item \textbf{Unregistered metric}: a metric handle cited but absent from the
ledger.
\item \textbf{Cross-section contradiction}: mutually inconsistent statements
across sections.
\end{enumerate}

\section{Ontology schemas}
\label{app:schema}
Both tiers speak the same three objects, which is why the bridge is a data adapter
rather than a rewrite. Required fields (from the JSON-Schema definitions the two
tiers share):
\begin{itemize}[leftmargin=1.4em,itemsep=1pt]
\raggedright
\item \textbf{EvidenceCard}: \texttt{evidence\_id}, \texttt{project\_id},
\texttt{source\_id}, \texttt{fact}, \texttt{source\_tier}, \texttt{confidence};
\texttt{quote} required when \texttt{allowed\_use}${=}$\texttt{hard\_evidence}.
\item \textbf{MetricLedger row}: \texttt{metric\_id}, \texttt{company},
\texttt{metric}, \texttt{authoritative\_value}, \texttt{value\_norm},
\texttt{source\_tier}, \texttt{basis\_evidence\_id}, \texttt{as\_of},
\texttt{alternatives}.
\item \textbf{Claim}: \texttt{claim\_id}, \texttt{subject}, \texttt{predicate},
\texttt{object}, \texttt{supporting\_evidence},
\texttt{contradicting\_evidence}.
\end{itemize}

\section{The distributed writer, illustrated}
\label{app:distributed}
\begin{figure*}[t]
  \centering
  \includegraphics[width=\textwidth]{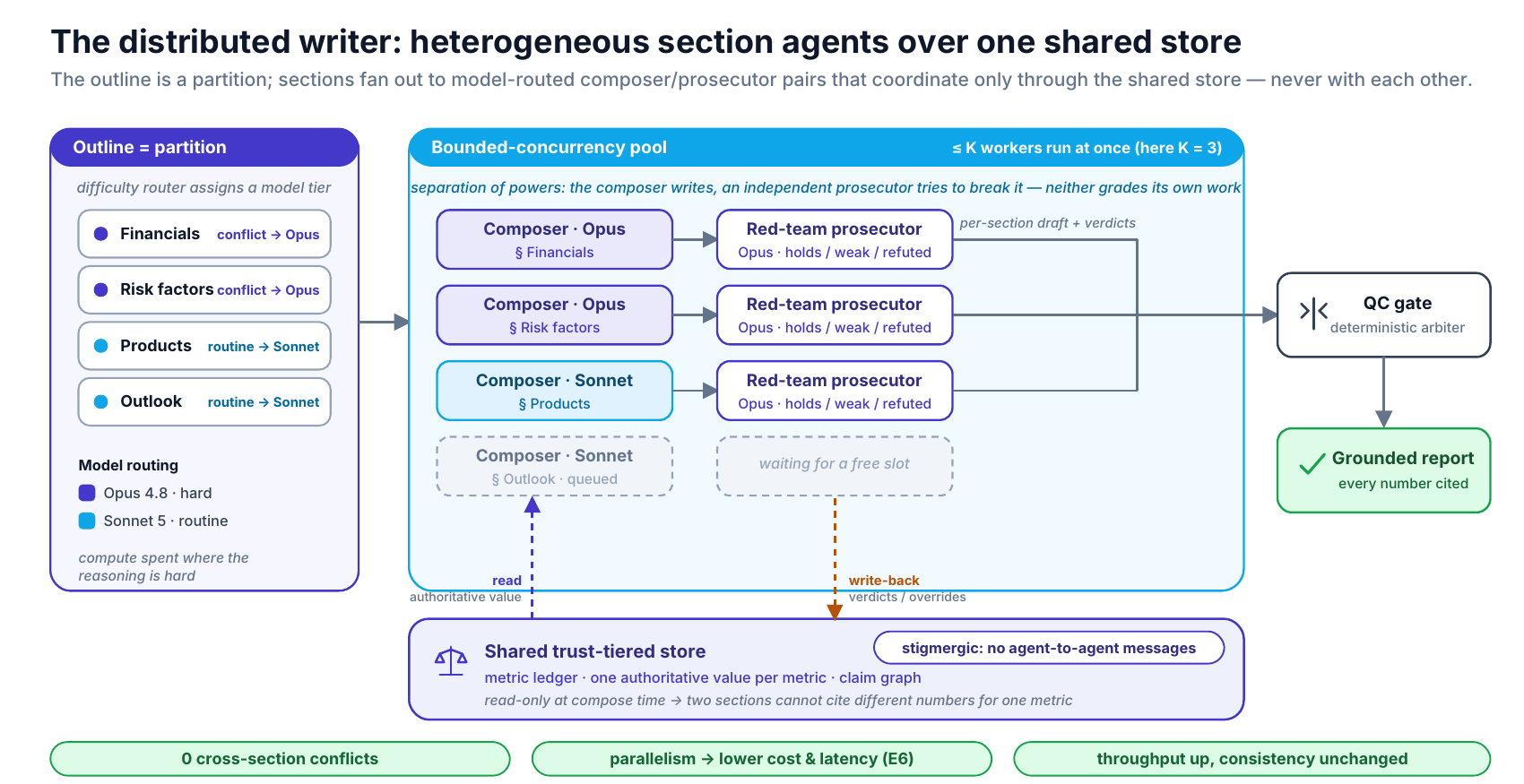}
  \caption{The distributed multi-agent writer (Phase~B, \S\ref{sec:distributed}).
  The report outline is partitioned into sections; a difficulty router sends
  conflict-touching sections to Claude Opus~4.8 and routine ones to Claude
  Sonnet~5. Sections fan out across a bounded-concurrency pool ($\le K$ workers),
  each a composer paired with an independent Opus red-team ``prosecutor'' that
  cannot grade its own work. Agents never message each other: they coordinate
  \emph{stigmergically} through the shared trust-tiered store (reading the single
  authoritative value per metric and writing verdicts back), so parallelism never
  costs cross-section consistency. A deterministic quality-control gate, not an LLM,
  is the final arbiter.}
  \label{fig:distributed}
\end{figure*}

Figure~\ref{fig:distributed} draws the four design points of \S\ref{sec:distributed}
as one picture: the outline is a partition whose sections a difficulty router assigns
to a model tier (heterogeneous agents); the sections fan out across a
bounded-concurrency pool of composer/red-team pairs (parallelism and separation of
powers); and the agents coordinate only through the shared trust-tiered store (read
at compose time, written back by the red-team) rather than by messaging each other
(stigmergic coordination). The single authoritative value lives in the store, so no
two sections can cite different numbers for the same metric even while they run
concurrently.

\section{Results at a glance}
\label{app:results}
\begin{figure*}[t]
  \centering
  \newcommand{\rescell}[2]{%
    \begin{minipage}[t]{0.32\textwidth}\centering
      \includegraphics[width=\linewidth]{#1}\\[1pt]
      {\footnotesize #2}
    \end{minipage}}
  \rescell{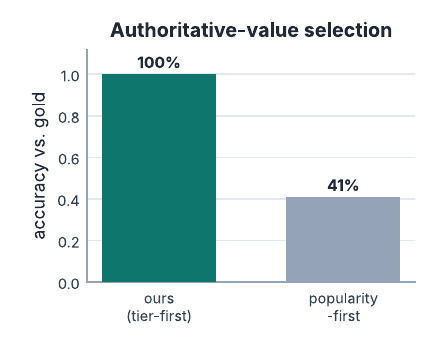}{(a) E4: selection accuracy vs.\ gold
    ($22/22$ vs.\ $9/22$)}\hfill
  \rescell{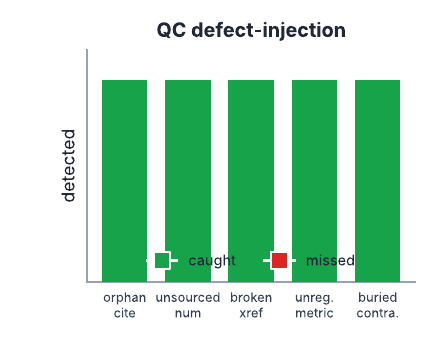}{(b) E5: QC recall \& precision
    ($1.00$ / $1.00$)}\hfill
  \rescell{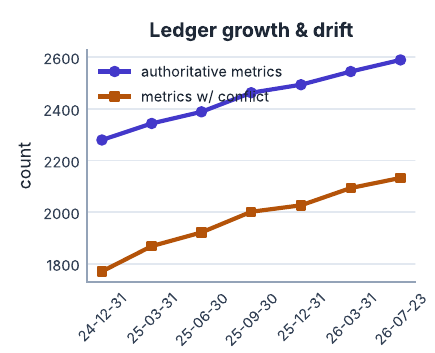}{(c) E1: ledger growth \& drift
    ($100\%$ of $4{,}732$ justified)}

  \vspace{7pt}
  \rescell{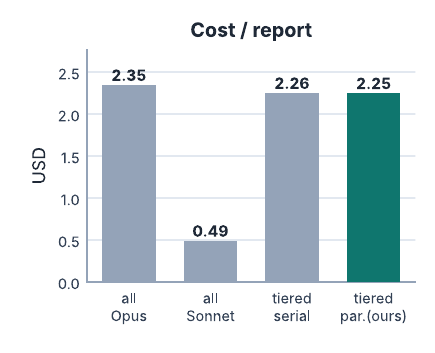}{(d) E6: cost / report (USD)}\hfill
  \rescell{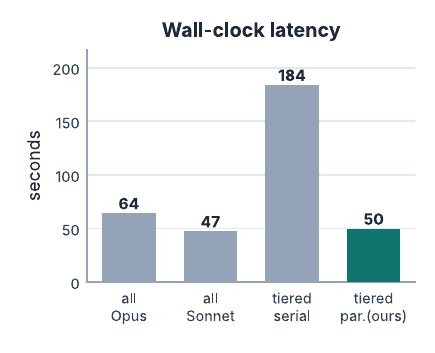}{(e) E6: wall-clock latency (s)}\hfill
  \rescell{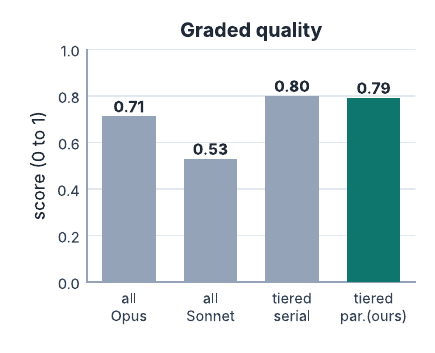}{(f) E6: graded quality ($0$--$1$)}
  \caption{Headline results as vector charts (companion to
  Tables~\ref{tab:e1}--\ref{tab:e6}), on a $2\times3$ grid: top row, the
  consistency and evaluation results (E4/E5/E1); bottom row, the E6
  cost/latency/quality trade-off across routing variants (ours in teal: cost
  at the all-Opus level, $3.7\times$ faster than serial, quality above the
  ceiling).}
  \label{fig:results}
\end{figure*}

Figure~\ref{fig:results} visualizes the headline quantitative results whose
exact figures are tabulated in \S\ref{sec:eval}, on a $2\times3$ grid: tier-first
vs.\ popularity-first selection (E4), QC defect-injection recall and precision
(E5), ledger growth with fully-justified drift (E1), and the E6 cost, latency, and
graded-quality trade-off across routing variants.

\section{Write-back trace}
\label{app:tables}
Table~\ref{tab:e8} traces the one worked write-back correction of E8
(\S\ref{sec:eval}) step by step: a red-team refutation maps to a librarian source
override that self-corrects the ledger's authoritative value at the next cutoff.

\begin{table}[h]
  \centering
  \caption{E8: a red-team refutation writes back and self-corrects the ledger.}
  \label{tab:e8}
  \small
  
\begin{tabular}{@{}l>{\raggedright\arraybackslash}p{0.52\columnwidth}@{}}
\toprule
Write-back step & state \\
\midrule
Metric (worked case) & issuer $X$, interest \\
Authoritative value (before) & \$19mn \\
Red-team verdict & \texttt{refuted} on backing filing \\
Librarian action & \texttt{source\_override}: retract; 284 cards re-stamped \\
Authoritative value (after) & \textbf{\$9mn} (prior alternative) \\
Metrics self-corrected (batch) & 5 of 6 traced \\
Manual value edits (all cases) & 0 \\
\bottomrule
\end{tabular}

\end{table}

\section{Reproducibility}
\label{app:repro}
Deterministic experiments (E1, E3, E4, E5, E7, E8) run with no cloud and reproduce
exactly from a fixed store; E6 hits the real Bedrock backend. Every table and figure
is emitted programmatically from the stored per-experiment result records. The
deterministic experiments run against an offline stub backend, and each source's
per-tier licensing and fetch parameters are recorded with the stored corpus.

\section{Ethical, legal, and societal considerations}
\label{app:ethics}
\begin{sloppypar}
The system operates only on public, timestamped financial sources used under each
source's terms (with per-tier licensing documented): it ingests and
reports metrics but never emits any customer, personal, or otherwise non-public
data, and every delivered figure is traceable to its source, so provenance is
auditable rather than obscured.
\end{sloppypar}

\end{document}